\documentclass[12pt]{article}

\newcommand{\bbr}{I\!\! R}

\newcommand{\x}{arXiv:}
\newcommand{\m}{\mathrm}

\usepackage{graphicx}

\begin{document}
\thispagestyle{empty}
\begin{center}

\null \vskip-1truecm \vskip2truecm {\Large{\bf

\textsf{Inaccessible Singularities in Toral Cosmology}

}}

\vskip1truecm {\large \textsf{Brett McInnes}} \vskip1truecm

\textsf{\\  National
  University of Singapore}

email: matmcinn@nus.edu.sg\\

\end{center}
\vskip1truecm \centerline{\textsf{ABSTRACT}} \baselineskip=15pt
\medskip
The familiar Bang/Crunch singularities of classical cosmology have
recently been augmented by new varieties: rips, sudden
singularities, and so on. These tend to be associated with final
states. Here we consider an alternative possibility for the initial
state: a singularity which has the novel property of being
\emph{inaccessible} to physically well-defined probes. These
singularities arise naturally in cosmologies with toral spatial
sections.

\newpage

\addtocounter{section}{1}
\section* {\large{\textsf{1. Singularities and Winding}}}
The study of singularities is potentially a powerful tool for
understanding quantum cosmology. Even if a given kind of singularity
is not expected to arise in our Universe, a close examination may
reveal much about the nature and structure of the gravitational
theory in the background. This is particularly true if one hopes
that that theory \emph{resolves} cosmological singularities, since
the exact meaning of ``resolving" a singularity is still not
entirely clear in all cases.

Recently, new classes of singularities have come under scrutiny: in
particular, new kinds of \emph{future} singularity have been studied
extensively \cite{kn:sergei}\cite{kn:copeland}, including rips,
sudden singularities, and so on. Some of these have very unusual
properties \cite{kn:lazkoz} which may well throw light on the
meaning of singularity resolution. Here we shall consider a kind of
\emph{past} singularity with novel properties that may allow a
better understanding of cosmological spacetimes in which extended
objects [strings, branes] play an important role at early times.

String theory has suggested various ways of dealing with initial
cosmological singularities \cite{kn:tsuji}. Among the most promising
ideas in this direction is that of applying T-duality to toral
spatial dimensions in cosmological models. The well-known
Brandenberger-Vafa string gas cosmologies \cite{kn:bat} make use of
T-duality to argue that the initial singularity cannot really be
present, since very small toral spatial sections must be physically
equivalent to very large, and hence non-singular, sections. The
toral topology is directly relevant, since of course it permits
\emph{string windings}. This possibility is also fundamental to the
[quite different] approach to singularity resolution explored in
\cite{kn:end}\cite{kn:silv}. Finally, we have argued
\cite{kn:OVV}\cite{kn:arrow} that toral sections play a crucial role
in the string-theoretic version \cite{kn:ooguri} of ``creation from
nothing", which also solves the singularity problem.

There is a mystery here, however, because, under the inflationary
conditions which apparently obtained in the very early Universe,
\emph{toral spatial sections actually make singularities harder to
avoid.} This is a general way of stating certain very remarkable
results due to Andersson and Galloway
\cite{kn:andergall}\cite{kn:gall}.

In more detail, the relevant theorem can be phrased in physical
language as follows. Consider a four-dimensional Lorentzian manifold
M, which is intended to represent the structure of spacetime up to
the end of an assumed inflationary era. We can therefore suppose
that M is generically future asymptotically de Sitter\footnote{That
is, it has a regular future spacelike conformal boundary; see
\cite{kn:andergall}\cite{kn:singularstable} for definitions.} and
non-singular to the future\footnote{That is, it is future
asymptotically simple.}. Suppose that the Null Ricci Condition [see
below] holds, and that the spatial sections of spacetime at the end
of Inflation have the topology of a torus\footnote{More generally,
we just have to assume that the future conformal boundary has a
covering by a compact orientable manifold with a first homology
group containing at least one element of infinite order.}.
\emph{Then M cannot be past null geodesically complete.}

The \emph{Null Ricci Condition} requires that the Ricci tensor
should satisfy, at each spacetime point,
\begin{equation}\label{eq:NULLRICCI}
\m{R_{\mu\nu}\,k^\mu\,k}^\nu\;\geq\;0
\end{equation}
for every \emph{null} vector k$^\mu$. This condition is not an
energy condition; it is purely geometrical. \emph{If} the Einstein
equations hold at all times, it is equivalent to the familiar Null
Energy Condition; see \cite{kn:arrow} for a discussion of this.

The Andersson-Galloway is a singularity theorem, and a very powerful
one, since the conditions required for it to apply are extremely
weak. Notice particularly that, apart from the Null Ricci Condition,
all of the conditions are imposed at the \emph{end} of Inflation
---$\,$ none of them is applied to the pre-inflationary era, about
which very little is known.

Throughout this work we shall assume that the Universe did pass
through an inflationary phase. The theorem then leaves us with only
two options for toral cosmology: either we violate the Null Ricci
Condition, or we accept that the singularity is really present, and
try to tame it in some other way\footnote{The theorem can of course
be circumvented by abandoning Inflation: see \cite{kn:brand}. It
does apply, however, to any inflationary model, with or without
\cite{kn:scott} an inflaton.}.

The first option has been explored in
\cite{kn:unstable}\cite{kn:tallandthin}\cite{kn:singularstable}\cite{kn:arrow}.
We believe that it is a promising approach, but one must bear in
mind that it faces its own difficulties. In particular, we should
stress that extended objects, such as branes, can be directly
sensitive to the ambient spacetime geometry, which controls the rate
at which their areas and volumes change as they are moved. It has
been found that severe non-perturbative instabilities can arise as a
response to these geometric conditions \cite{kn:porrati}. In string
cosmology, it turns out
\cite{kn:unstable}\cite{kn:tallandthin}\cite{kn:singularstable} that
spacetime geometries violating the NRC \emph{generically} lead to
this kind of instability, under the assumption that there should be
no propagation of energy outside the null cone. There are
interesting, very special exceptions, and we believe that these
exceptions might yield an accurate picture of the spacetime geometry
of the early Universe \cite{kn:arrow}.

In view of the delicacy of the argument, however, it is as well to
consider the second possibility. In this work, we shall
\emph{maintain} the NRC, and examine the nature of the singularity
which must then be present in the pre-inflationary era. We shall
argue that, while a singularity is inevitable in this case, the
singularity could well be of a novel kind: it could be
\emph{inaccessible}. The incomplete region cannot actually be probed
by real physical objects with well-defined properties. By this we
mean that a generic timelike or null geodesic can only reach the
singularity by wrapping around the torus \emph{infinitely many
times} in a finite amount of proper time; and we shall see that this
is a paradoxical situation when discrete symmetries are taken into
account. In this sense, the singularity, while still present in
idealized mathematical models, is unphysical.

The simplest spacetime to which the Andersson-Galloway theorem
applies is the spatially toral version of de Sitter spacetime
itself. Let us begin by examining this spacetime in detail.

\addtocounter{section}{1}
\section*{\large{\textsf{2. The Ultimate Twins}}}
Consider the familiar version of de Sitter spacetime with flat
spatial sections, with spacetime length scale L [associated with the
cosmological constant] and with metric
\begin{equation}\label{eq:C}
g(\m{SCdS)_L \;=\; dt^2\; -\;e^{(2\,t/L)}\,[\,dx^2 \;+\; dy^2 \;+\;
dz^2}\,].
\end{equation}
We may call this ``Spatially Cartesian de Sitter spacetime". In
order to allow for string winding states, let us compactify the
spatial sections to cubic tori
--- the details of this will be discussed below --- so that the
metric becomes
\begin{equation}\label{eq:STDS}
g(\m{STdS)(\,K,\,L)}\;=\;\m{dt^2\;
-\;K^2\,e^{(2\,t/L)}\,[\,d\theta_1^2 \;+\; d\theta_2^2 \;+\;
d\theta_3^2]},
\end{equation}
where we are using angular coordinates on the circles of which the
torus is comprised and K is the radius of these circles at t = 0.

The Andersson-Galloway theorem demands that this spacetime should
be past null geodesically incomplete. For clearly the spacetime is
asymptotically de Sitter, with a toral conformal boundary, the NRC
is satisfied, and so on. Let us explain how this can be.

We begin by noting that ``Spatially Toral Minkowski" spacetime, with
metric
\begin{equation}\label{eq:STM}
g(\m{STM)_K}\;=\;\m{dt^2\; -\;K^2\,[\,d\theta_1^2 \;+\; d\theta_2^2
\;+\; d\theta_3^2]},
\end{equation}
is an interesting object in its own right. It has sometimes been
considered in connection with the ``Topological Twins", the variant
of the traditional relativistic twins in which the toral topology
allows the twins to meet again \emph{without} either needing to
accelerate. [See \cite{kn:roukema} for a recent discussion with many
references.] The twins still have different ages when they meet,
despite both having geodesic worldlines. The ``paradox" is resolved
by noting that the topological identifications which allow the twins
to meet have broken the spacetime isometry group down to a subgroup
which is too small to map [even segments of] one geodesic onto the
other.

We shall now study a pair of such twins in the STdS spacetime
geometry.

As is well known [see for example \cite{kn:tallandthin}], the
Cartesian coordinates used in equation (\ref{eq:C}) cover only
\emph{half} of de Sitter spacetime, the ``upper triangle" in the
familiar square Penrose diagram of the full [topology
$\bbr\,\times\,\m{S}^3$] de Sitter spacetime dS$_4$. Physically,
this is precisely the set of all events in dS$_4$ to which a signal
can be sent by an inertial observer at the origin of Cartesian
coordinates. This is in fact an open submanifold of
$\bbr\,\times\,\m{S}^3$: the Penrose diagram is obtained by deleting
from the usual square diagram the lower triangular half,
\emph{including the diagonal and its endpoints}. Thus, the harmless
appearance of (\ref{eq:C}) is deceptive: the spacetime is
pathological in the sense that objects can suddenly enter it from
``outside". In fact, the harmless appearance of the Cartesian de
Sitter metric arises simply because we choose to describe the
geometry using a family of inertial observers such that all of the
corresponding spatial sections just barely avoid the region which
has been cut away.

\begin{figure}[!h]
\centering
\includegraphics[width=0.6\textwidth]{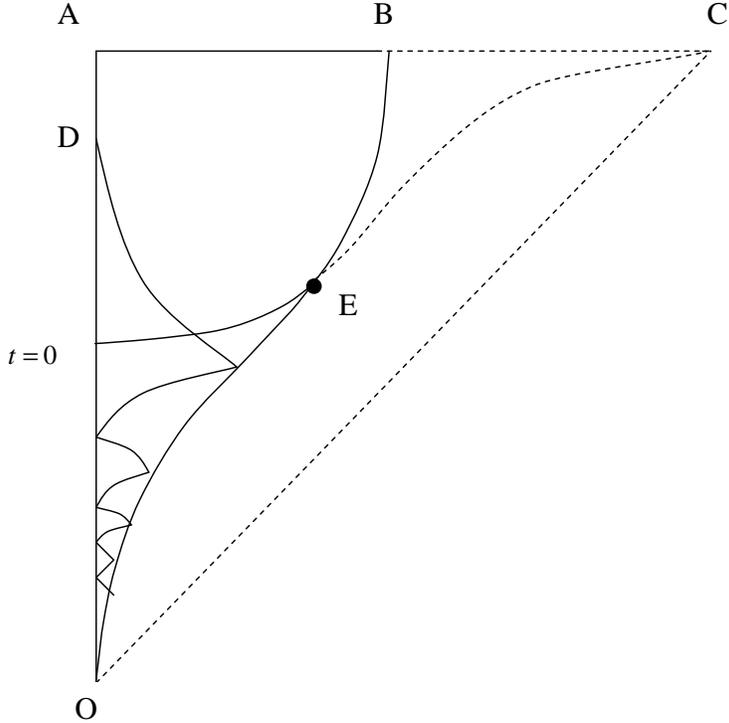}
\caption{Penrose Diagram for Spatially Toral de Sitter Spacetime.}
\end{figure}

The timelike curve OEB shown in Figure 1 represents the worldline of
one of these observers, one who passes through the event E
corresponding to the two-sphere of radius $\pi$K in the spacelike
surface t = 0 [in which the Cartesian coordinates x,y,z measure
proper length]. This two-sphere can be enclosed in a cube of side
length 2$\pi$K, and this cube can be allowed to expand or contract
as we follow the sphere either into the future or the past along the
chosen geodesic. We obtain Spatially Toral de Sitter [STdS]
spacetime by performing the usual topological identifications of the
opposite faces of these cubes. [We remind the reader that more
complicated identifications can be performed if one so desires; see
\cite{kn:arrow} for references.] The effect on the Penrose diagram
is [to a good approximation] to cut away all parts of the original
triangle which lie to the right of the chosen geodesic OEB; this is
the dotted region in Figure 1. Note that conformal infinity is also
compactified to a torus: it is represented by the solid part, AB, of
the upper horizontal line.

It is clear that, with the crucial exception of the point O in
Figure 1, no part of the STdS Penrose diagram intersects the line OC
along which the original [uncompactified] spacetime was geodesically
incomplete. Furthermore, all of the geodesics perpendicular to the
surfaces t = constant are inextensible, since the coordinate t
measures proper time for them, and the point O is at ``t =
$-\,\infty$". These facts may explain the widespread belief that
Spatially Toral de Sitter spacetime is geodesically complete.
\emph{This is not correct}, however, as Galloway \cite{kn:gall} has
emphasised. The point is that the timelike geodesics defining the
coordinates used in equation (\ref{eq:STDS}) do not pass through the
faces of the distinguished cube, because they are ``co-moving" with
that cube. Other timelike and null curves \emph{will} do so,
however, since they represent objects with non-zero velocities
relative to the fundamental observers; and of course one has to
check whether the proper time [or affine parameter] along \emph{all}
of these curves, too, is infinite. In fact, the Andersson-Galloway
theory implies that it is not, and we can verify this explicitly as
follows.

Consider the timelike geodesics G$_1$, G$_2$ shown in Figure 1,
where G$_1$ is just the t-axis OA, and G$_2$ is the curve extending
down from D which appears to ``bounce" back and forth between the
origin and the distinguished geodesic OEB. [Being timelike, this
geodesic should not be represented by straight segments in the
conformal diagram, and we have made an attempt to show this.] In
reality, G$_2$ is smooth: it is just winding around the torus. [The
``bounces" on the left side of the diagram represent the curve
passing through the origin and continuing through to the other side;
the ``bounces" on the right represent what happens when the curve
passes through a face of the distinguished cube.] If we trace it
back to the point O, it will wrap around the torus \emph{infinitely}
many times.

In the metric (\ref{eq:STDS}), the vector $\partial/\partial\m{t}$
is not a Killing vector, but $\partial/\partial\theta_i$, i = 1,2,3,
clearly is; thus, the product of this field with the tangent vector
to a timelike geodesic [parametrized by proper time] is constant
along the geodesic. We have, for any timelike geodesic,
\begin{equation}\label{eq:D}
\m{K^2\,e^{(2t/L)}\,{{d\theta_i}\over{d\tau}}\;=\;s_i},
\end{equation}
where s$_\m{i}$ is a constant along the geodesic and $\tau$ is
proper time along this geodesic. One can think of s$_\m{i}$ as a
parameter which distinguishes the speeds [in the $\theta_{\m{i}}$
direction] of freely falling objects, relative to the local observer
with constant spatial coordinates, at [say] t = 0. Clearly the
s$_\m{i}$ are all zero if and only if the geodesic is one of those
corresponding to the t-coordinate lines [such as G$_1$]. Now since
the tangent is a unit vector, we have, assuming that the object is
moving in the $\theta_1$ direction,
\begin{equation}\label{eq:E}
\m{\Big({{dt}\over{d\tau}}\Big)^2\;=\;1\;+\;K^2\,e^{(2t/L)}\,\Big({{d\theta_1}\over{d\tau}}\Big)^2},
\end{equation}
and so
\begin{equation}\label{eq:F}
\m{\Big({{dt}\over{d\tau}}\Big)^2\;=\;1\;+\;{{s_1^2}\over{K^2}}\,e^{(-\,2t/L)}}.
\end{equation}
If t = T at the point D in Figure 1, then the total proper time
P(G$_{1,2}$) along either G$_1$ or G$_2$ is
\begin{equation}\label{eq:G}
\m{P(G_{1,2})\;=\;\
\int_{-\,\infty}^T{{dt}\over{\sqrt{1\;+\;s_1^2\,K^{-\,2}\,e^{-\,2t/L}}}}}\;.
\end{equation}

Now G$_1$ and G$_2$ are supposedly ``twins": they enter the
spacetime at O. What is the difference in their ages at D? For
G$_1$, s$_1$ is exactly zero, so G$_1$ is \emph{infinitely} old at
D. For G$_2$, by contrast, we have
\begin{equation}\label{eq:H}
\m{P(G_2)\;=\;L\,sinh^{-\,1}\Big({{K}\over{s_1}}\,e^{T/L}\Big)},
\end{equation}
which is finite since s$_1$ is not zero here. Thus we have the most
extreme version of the ``paradox of the twins": one of the twins is
not merely older, as in Spatially Toral Minkowski spacetime --- he
is in fact \emph{infinitely} older than the other. This infinite
difference is clearly due to the fact that G$_2$'s worldline wraps
around the torus infinitely many times. The finite age of G$_2$
means that spatially Toral de Sitter spacetime is indeed timelike
geodesically \emph{incomplete}, since it contains inextensible
timelike geodesics of finite length \cite{kn:gall}.

For a null geodesic N extending down from D, representing a ray of
light entering the spacetime at O and propagating in the $\theta_1$
direction, a similar calculation [identical except that 0 replaces 1
on the right side of (\ref{eq:E}), and affine parameter replaces
proper time] yields
\begin{equation}\label{eq:I}
\m{A(N)\;=\;L\,{{K}\over{s_1}}\,e^{T/L}},
\end{equation}
where A(N) is the total affine parameter along N; this is always
finite since s$_1$ cannot be zero in this case. Thus the spacetime
is also \emph{null} geodesically incomplete, as the
Andersson-Galloway theorem requires.

The mathematics of this situation is elementary, but its real
physical meaning is extremely obscure. The question as to whether
this Universe has a beginning is a matter of opinion: one will get
different answers from G$_1$ and G$_2$. Even among observers who
agree that the Universe does have a beginning, there will be
arbitrarily large disagreements as to how old it is. For the right
side of equation (\ref{eq:H}) diverges as s$_1$ tends to zero, so
the slightest difference between two small values of s$_1$ leads to
vast differences between the respective ages of the corresponding
objects. Thus the very concept of ``the age of the Universe" becomes
incoherent.

Notice that actually ``reaching" the point O [that is, following
G$_2$'s finite history back to his ``beginning"] is non-trivial, in
the following physical sense: if G$_2$ has three-momentum p$_0$
relative to G$_1$ at t = 0, then his momentum at other times is just
\begin{equation}\label{eq:I}
\m{p\;=\;p_0\,e^{-\,t/L}};
\end{equation}
that is, it is infinitely ``blue-shifted" as we trace G$_2$'s
history back to O, which is at t = $-\,\infty$. His energy is
similarly blue-shifted. Thus, the point O can only be reached at
the cost of an infinite growth of the probe's momentum and energy.
This cannot be blamed on the presence of a curvature singularity,
since the local geometry of this spacetime is everywhere that of
de Sitter spacetime. \emph{It is, once again, related to the fact
that O can only be reached in a finite proper time by winding
around the torus an infinite number of times.}

Very strange consequences follow from the fact that all freely
falling objects which circumnavigate this Universe do so a
\emph{literally infinite} number of times. Suppose, for example,
that the spatial sections are not tori but rather the product of a
Klein bottle with a circle, KB$^2\,\times\,$S$^1$, with the flat
three-dimensional metric. Each circumnavigation around the
``twisted" direction of the Klein bottle reverses orientation. At
the point D in Figure 1, by which time G$_2$ --- the twin whose age
is \emph{finite} --- has had his orientation reversed infinitely
many times, does G$_2$ still have the same ``handedness" as G$_1$?
There is no answer to this question. Again, it is possible
\cite{kn:preskill} for parallel transport around non-contractible
loops to convert particles to anti-particles; once again, this means
that we simply cannot say whether the ``twins" are both made of
particles or of antiparticles when they meet. Similar ambiguities
can arise for time orientation. \emph{Such ``discrete holonomy
effects" are generic for topologically non-trivial spacetimes;} and
yet they clearly do not make sense when the spacetime structure
leads to infinite numbers of circumnavigations in finite amounts of
proper time. \emph{The incomplete region is inaccessible to objects
with well-defined discrete [generalized] ``orientations".}

There is an important example where parallel transport \emph{does}
generate a discrete transformation. On a torus, T$^3$, the parallel
transport of vectors is trivial. However, the parallel transport of
spinors around T$^3$ need not be trivial: since the spin holonomy
group of any spin manifold projects onto its vector holonomy group,
the sign of a spinor can be changed by parallel transport around a
non-contractible loop. Spin connections of this kind are an
essential ingredient of many constructions in string cosmology: for
example, it is explicitly required in reference \cite{kn:end} [in
the guise of ``antiperiodic fermion boundary conditions"], so in
winding string tachyon theory we are precisely in the situation
described above. The sign of a spinor changes infinitely many times
in a finite interval of proper time.

\addtocounter{section}{1}
\section*{\large{\textsf{3. Extended Objects}}}
The discussion thus far concerns [toral] de Sitter spacetime,
which of course is not singular [in the sense that its curvature
never diverges, even in the incomplete region]; furthermore, its
only matter content is that represented by the cosmological
constant, which we can think of as the inflaton in its
``potential-dominated" state. In toral cosmology, however, it is
natural to suppose that, in addition to the inflaton, the earliest
Universe contained winding strings and other extended objects,
which will deform the geometry and possibly lead to a true
singularity. Is this singularity, too, ``inaccessible"?

Let us introduce into toral de Sitter spacetime a static,
three-dimensional network of straight strings. [Note that with
toral topology, straight strings can be winding strings.] Thinking
of this network as a ``fluid", one easily sees that the energy
density and pressure are related by an equation of state given by
\begin{equation}\label{eq:GG}
\m{p_{string}\;=\;{{-\,1}\over{3}}\,\rho_{string}}.
\end{equation}
The Einstein equations can be solved in this case, in the
following interesting way. Since
$\m{\rho_{string}\;+\;3\,p_{string}\;=\;0}$, the contribution of
the strings to the right side of the Raychaudhuri equation
\begin{equation}\label{eq:RAY}
\m{3\,\Big({{\ddot{a}}\over{a}}\Big)\;=\;-\,4\,\pi\,(\,\rho\:+\;3\,p\,)}
\end{equation}
is \emph{exactly zero}, and so the only non-zero contribution is
made by the background cosmological constant; so the equation is
just $\m{\ddot{a}/a\;=\;1/L^2}$. This equation essentially has three
kinds of solution, involving cosh(t/L), exp(t/L), or sinh(t/L). But
the NRC here just means that the Hubble parameter $\m{\dot{a}/a}$
cannot increase; so leaving aside exp(t/L) [which is the case
discussed in the previous section, that is, it corresponds to
$\m{p_{string}\;=\;\rho_{string}}$ = 0], we see that the NRC [which
is certainly satisfied here, since
$\m{\rho_{string}\;+\;p_{string}\;>\;0}$] forces us to take the
sinh(t/L) solution. The metric is
\begin{equation}\label{eq:IOTAIOTA}
g\m{_{string}(\,K,\,L) \;=\; dt^2\; -\; K^2\;sinh^2
\Big({{t}\over{L}}\Big)\,[d\theta_1^2 \;+\; d\theta_2^2 \;+\;
d\theta_3^2].}
\end{equation}

Now this metric is geodesically incomplete at t = 0; it has a
curvature singularity there. We see that a mixture of toral spatial
sections and winding strings does \emph{not} eliminate the initial
singularity. Notice that this cannot be understood by means of the
classical singularity theorems, since the Strong Energy Condition is
actually violated \emph{at all times} in this spacetime. It can only
be understood by using the Andersson-Galloway theorem.

However, the singularity here arises in a peculiar way. Normally one
thinks of non-relativistic matter or radiation as being the ``cause"
of the singularities in the standard FRW models, in the sense that
they contribute to the right side of the Raychaudhuri equation in
such a way that spacetime geodesics are compelled to focus at a
caustic. But we saw that the strings \emph{make no contribution} to
the right side of the Raychaudhuri equation --- they have no role to
play in focussing geodesics as we move back in time. The strings do
not ``cause" the singularity in this direct sense: their influence
is felt only through the \emph{integral} of the Raychaudhuri
equation, which of course determines the scale factor. This suggests
that there must be something very unusual about the physical nature
of the singularity here.

In fact, as in the case of pure Spatially Toral de Sitter spacetime,
the singularity here is inaccessible. To see this, we merely need to
note that the expression for total conformal time,
\begin{equation}\label{eq:ITHACA}
\m{H(L/K,\;2)\;=\;{{L}\over{K}}\,\int_0^{\infty}\,{{dx}\over{sinh(x)}}},
\end{equation}
is clearly divergent. [See \cite{kn:sbound} for the notation and a
more general discussion of the behaviour of conformal time in
spacetimes with toral spatial sections.] It follows that the Penrose
diagram for the spacetime with metric $g\m{_{string}(\,K,\,L)}$ has
precisely the same shape as the one portrayed in Figure 1, so it is
still the case that objects which [by reversing the direction of
time] we might regard as ``probing" the singularity will in general
have to perform an infinite number of circumnavigations in order to
do so. In fact, this will always be the case if the earliest history
of the Universe was dominated by some kind of extended object ---
strings or any kind of brane. Again, it does not make sense [for
example] for a fermion to behave in this way when, as is the case in
the winding tachyon scenario, one has imposed antiperiodic boundary
conditions.

We conclude that, whatever may be the case mathematically, it is
physically not meaningful to probe these spacetimes back to ``the
beginning of time". If we are to picture the period when string
winding dominated the dynamics, we should think of it as a chaotic
era in which a particle's age and spin orientation, and perhaps its
charge and other generalized ``orientations", are so sensitively
dependent on earlier conditions that they lose all meaning. The
singularity is still technically present, as the Andersson-Galloway
theorem demands, but it inhabits a region of spacetime which cannot
be probed by objects with well-defined properties.

 \addtocounter{section}{1}
\section*{\large{\textsf{4. Conclusion}}}

One of the fundamental features of Inflation is its tendency to
eradicate evidence of whatever preceded it. This is of course often
desirable, but it apparently does not help us to understand the
initial singularity. In this work we have seen an example where
Inflation \emph{does} tell us something important about what
preceded it: in cases where the spatial sections at the end of
Inflation have at least one circular factor, the existence of an
inflationary era actually greatly \emph{strengthens} the classical
cosmological singularity theorem. For the Andersson-Galloway theorem
allows a great weakening of the assumptions of the classical
theorem, from the Strong Energy Condition down to the Null Ricci
Condition. If the latter is not violated, it follows that spacetime
has to be singular \emph{precisely in the case of interest in string
cosmology}, that is, the case where toral topology permits closed
strings to wind.

We suggested that one possible response to this situation is to
accept the mathematical conclusion of the Andersson-Galloway
theorem, but to argue that it demands the presence of a singularity
which is \emph{physically inaccessible}. In simple models of string
winding spacetimes, it seems that the initial singularity cannot be
probed by physically well-defined objects. The earliest era of
string winding is pictured as a ``chaotic" phase in which the
spacetime structure forces the discrete properties of particles to
be so dependent on position in space and time as to be essentially
meaningless. This gives us a concrete realization of the intuitive
idea that space and time may lose their familiar physical
interpretations in the earliest Universe.

\addtocounter{section}{1}
\section*{\textsf{Acknowledgements}}
The author is grateful for very helpful correspondence from
Professor Galloway, and to Wanmei for the diagram.

\end{document}